\documentclass[conference,a4paper]{IEEEtran}
\IEEEoverridecommandlockouts
\usepackage{amsmath,epsfig,amssymb,verbatim}
\usepackage{cite}

\usepackage{graphicx}
\usepackage{bm,bbm}
\usepackage{soul}
\usepackage{xcolor}

\usepackage{textcomp}
\usepackage{caption}
\usepackage{subcaption}
\usepackage{float}
\usepackage{stackengine}
\usepackage{cleveref}
\usepackage{multirow}

\newtheorem{Theorem}{Theorem}

\newtheorem{Proposition}{Proposition}

\def\BibTeX{{\rm B\kern-.05em{\sc i\kern-.025em b}\kern-.08em
   T\kern-.1667em\lower.7ex\hbox{E}\kern-.125emX}}

\usepackage{mathtools} % improves multiline equations
\usepackage{dblfloatfix}
\sethlcolor{yellow}

\soulregister\ref7
\soulregister\eqref7
\soulregister\cite7
\soulregister\pageref7
\soulregister\textit7
\soulregister\emph7
\soulregister\mathbf7
\soulregister\mathrm7
\soulregister\mathcal7

\usepackage{caption}
\begin{document}

\title{QoE-Driven Hybrid Resource Allocation for Multi-User OFDMA Semantic Communications}
\author{\IEEEauthorblockN{Xiaotian Xu$^{\dagger}$, Zhifeng Tang$^{\dagger}$, Guangchen Wang$^{\dagger}$, Nan Yang$^{\dagger}$, Xiangyun Zhou$^{\dagger}$,}
\IEEEauthorblockA{ Salman Durrani$^{\dagger}$, and Yansha Deng$^{\ddagger}$}
\IEEEauthorblockA{$^{\dagger}$School of Engineering, Australian National University, Canberra, ACT 2600, Australia}
\IEEEauthorblockA{$^{\ddagger}$Department of Engineering, King’s College London, London WC2R 2LS, United Kingdom}
\IEEEauthorblockA{Emails: \{xiaotian.xu, zhifeng.tang, guangchen.wang, nan.yang, xiangyun.zhou, salman.durrani\}@anu.edu.au,}
\IEEEauthorblockA{yansha.deng@kcl.ac.uk}}

\maketitle

\begin{abstract}
In this paper, we develop a quality-of-experience (QoE)-driven hybrid resource allocation framework for multi-user downlink orthogonal frequency-division multiple access semantic communication systems, where the hybrid framework jointly handles continuous and discrete decision variables. Specifically, we formulate a mixed continuous-discrete optimization problem to jointly optimize transmit power, subcarrier allocation, and semantic decisions under coupled power and spectrum constraints. To better characterize semantic service quality, we propose a preference-inverse multiplicative exponential QoE that is bounded and non-compensatory, capturing the intrinsic ``both-required'' property of semantic rate and similarity. To solve the optimization problem, we develop a projection-based multi-agent deep deterministic policy gradient framework that enforces feasibility for continuous and discrete actions. Furthermore, we integrate Gumbel-softmax relaxation with an adaptive performance-driven temperature annealing scheme to enhance the learning of discrete semantic decisions. Simulation results demonstrate that our proposed framework achieves superior convergence, greater stability, and higher QoE compared to the conventional QoE design and penalty-based or fixed-temperature baselines, highlighting the benefits of our QoE design and the effectiveness of hybrid decision learning in semantic communications.
\end{abstract}

\begin{IEEEkeywords}
Semantic communications, quality of experience (QoE), MADDPG, adaptive temperature annealing.
\end{IEEEkeywords}

% =========================

\section{Introduction}

Semantic communications has emerged as a transformative wireless paradigm by prioritizing task-relevant semantic information over bit-level fidelity, significantly improving communication efficiency in text-focused applications such as text delivery and intelligent interactions \cite{Ref1}. In practical wireless networks, however, semantic transmission performance depends on both semantic representation and radio resource allocation among multiple users. Heterogeneous semantic requirements, coupled power and spectrum constraints, and the interaction between communication and semantic processing make joint semantic processing and resource allocation essential \cite{Lu2024Tutorial}.

Motivated by the above, recent studies have investigated semantic-aware resource allocation through various lenses, including metric design, multi-task networks, and learning-based resource control \cite{SR,QoE,Ref6,RefNew1}. Early studies have mainly focused on text semantic communications and quality of experience (QoE)-aware utility design \cite{SR,QoE}, while later studies have shifted this focus to multi-task networks and learning-based control in dynamic environments \cite{Ref6,RefNew1}. Despite these advances, two fundamental challenges remain. First, existing studies have mainly formulated the QoE as a weighted sum of semantic rate and similarity \cite{QoE,GenGap}. This additive formulation is compensatory and unsuitable for semantic services necessitating the simultaneous fulfillment of both semantic rate and similarity. Second, existing differentiable action mappings have primarily been designed for single-type constraints \cite{DeepSC}, making them ill-suited for hybrid continuous-discrete decisions under coupled resource budgets \cite{Hung2025ACRL}. In practical semantic communications, the controller must jointly optimize continuous transmit power, discrete subcarrier allocation, and discrete semantic decisions under coupled power and spectrum constraints, complicating end-to-end learning. These limitations motivate a unified framework for QoE design and hybrid decision-making under coupled constraints.

%To address such challenges, we develop a QoE-driven hybrid resource allocation framework for multi-user downlink orthogonal frequency-division multiple access (OFDMA) semantic communication systems. Specifically, we propose a bounded and non-compensatory preference-inverse multiplicative exponential QoE, a projection-based multi-agent deep deterministic policy gradient (MADDPG) framework, and an adaptive performance-driven temperature annealing (APTA) scheme for discrete semantic decision learning. Simulation results show superior convergence and stability.

To address such challenges, we develop a QoE-driven hybrid resource allocation framework for multi-user downlink orthogonal frequency-division multiple access (OFDMA) semantic communication systems. First, we propose a preference-inverse multiplicative exponential QoE which is bounded and non-compensatory, ensuring that poor performance in either semantic rate or similarity cannot be fully compensated by high performance in the other.
% and capable of continuously interpolating between rate-oriented and similarity-oriented user preferences. 
Second, we develop a projection-based multi-agent deep deterministic policy gradient (MADDPG) framework to handle coupled continuous and discrete resource constraints, together with discrete semantic decisions. Third, we design the adaptive performance-driven temperature annealing (APTA) scheme to improve the training of discrete semantic branches by balancing exploration and discretization. Simulation results demonstrate that the proposed framework achieves superior convergence and stability, offering key insights into how QoE design, projection-based constraint handling, and adaptive temperature control jointly enhance learning efficiency.

\section{System Model}

We consider the downlink of an OFDMA semantic communication system, where one base station (BS) serves $N$ users indexed by $\mathcal{Q}=\{1,2,\ldots,N\}$, with $U_n$ denoting the $n$-th user. The users are randomly distributed within a ring-shaped cell region. The distance between the BS and $U_n$ is denoted by $r_n$, where $r_n$ follows a uniform distribution, i.e., $r_n \sim \mathcal{U}(r_{\min}, r_{\max})$. Here, $r_{\min}$ and $r_{\max}$ denote the minimum and maximum distances, respectively, specifying the user-distance range in the considered system, ensuring the received signal quality to remain in a practical operating range.

%\vspace{-0.3em}

\subsection{Semantic Communication Architecture}

The system employs a deep learning-based semantic communication (DeepSC) architecture for text transmission \cite{DeepSC}. The BS uses a semantic encoder, whose model is pre-trained either locally or at a cloud server, while users possess corresponding DeepSC receivers with the trained encoder-decoder parameters distributed prior to deployment. For each user, the BS processes the source sentence and generates its semantic representation for transmission. The source sentence for $U_n$ is represented as $\mathbf{s}_n=[w_{n,1}, w_{n,2}, \ldots, w_{n,L_n}]$, where $w_{n,l}$ denotes the $l$-th word and $L_n$ denotes the number of words in a sentence (or equivalently, sentence length). The semantic encoder maps $\mathbf{s}_n$ into a transmitted symbol sequence $\mathbf{x}_n=[x_{n,1}, x_{n,2}, \ldots, x_{n,V_n}]$, where $x_{n,v}$ is the $v$-th normalized complex baseband channel symbol for $U_n$, $v\in\{1,\ldots,V_n\}$, and $V_n$ is the number of generated symbols. We define $d_n = V_n/L_n$ as the average number of semantic symbols per word (or equivalently, word length) for $U_n$.

Let $S_{\mathrm{tot}}$ and $P_{\mathrm{tot}}$ represent the total orthogonal subcarriers and total transmit power at the BS, respectively. Each subcarrier has a bandwidth $W$ and supports a symbol rate of $W$. In the system, the orthogonal subcarriers and total transmit power are allocated among users. Thus, we denote $S_n$ and $p_n\in[p_{\min},p_{\max}]$ as the allocated number of subcarriers and transmit power for $U_n$, respectively, where $p_{\min}$ and $p_{\max}$ denote the minimum and maximum per-user power, respectively. These per-user bounds mitigate extreme power disparities among users, thereby promoting inter-user fairness.
%prevent overly extreme user-level power allocation and also help maintain fairness across users.}
To maximize transmission efficiency, the BS jointly optimizes the resource allocation and semantic communication parameters $\{p_n, S_n, d_n\}$ for all users at each decision timestep.

Based on the determined semantic and communication resources, the generated semantic symbol sequence is transmitted over the downlink channel using the allocated subcarriers. Thus, the achievable symbol transmission rate of $U_n$ is $S_n W$ and the corresponding transmission duration is $V_n/(S_n W)$. At the user side, $U_n$ employs its DeepSC receiver to recover the semantic symbols and reconstruct the conveyed meaning.

%\vspace{-0.3em}

\subsection{Channel Model}

The signal propagation in the system is characterized by distance-dependent large-scale path loss and multipath small-scale fading. Under this characterization and assumption that $p_n$ is uniformly distributed over $S_n$ allocated subcarriers, the received signal for $U_n$ is given by
\begin{equation}\label{eq:received_signal}
y_n=\sqrt{\frac{p_n}{S_n}\mathrm{PL}(r_n)}  h_n  x_n + z_n,
\end{equation}
where $\mathrm{PL}(r_n)=\left({c}/{4\pi f_c}\right)^{-2}\left({r_n}/{r_0}\right)^{-\alpha}$ represents the large-scale path loss between the BS and $U_n$, with $c$, $f_c$, $r_0=1$ m, and $\alpha$ denoting the speed of light, carrier frequency, reference distance, and path loss exponent, respectively. Moreover, $h_{n}$ in \eqref{eq:received_signal} denotes the small-scale fading coefficient for $U_n$, modeled as Rayleigh fading, such that $h_n \sim \mathcal{CN}(0,1)$ follows a complex Gaussian distribution with zero mean and unit variance. We assume a user-level block-fading model such that $h_n$ remains constant over the subcarriers allocated to $U_n$ within each decision timestep but varies across different timesteps \cite{channel}. Furthermore, in \eqref{eq:received_signal},
%The small-scale fading is modeled as Rayleigh fading, such that the fading coefficient for $U_n$, $h_n \sim \mathcal{CN}(0,1)$, follows a complex Gaussian distribution with zero mean and unit variance. \hl{We assume a user-level block-fading model such that $h_n$ remains constant over the subcarriers allocated to $U_n$ within each decision timestep and varies across timesteps.}
%Specifically, the large-scale path loss between the BS and $U_n$ is modeled as
%\begin{equation}
%\mathrm{PL}(r_n)=\left(\frac{c}{4\pi f_c}\right)^{-2}\left(\frac{r_n}{r_0}\right)^{-\alpha},
%\end{equation}
%where $c$, $f_c$, $r_0=1$ m, and $\alpha$ denote the speed of light, carrier frequency, reference distance, and path loss exponent, respectively. The small-scale fading is modeled as Rayleigh fading, such that the fading coefficient for $U_n$, $h_n \sim \mathcal{CN}(0,1)$, follows a complex Gaussian distribution with zero mean and unit variance. \hl{We assume a user-level block-fading model such that $h_n$ remains constant over the subcarriers allocated to $U_n$ within each decision timestep and varies across timesteps.}
%For $U_n$, the transmit power $p_n$ is uniformly distributed over the $S_n$ allocated subcarriers. Thus, the received signal is
%\begin{equation}
%y_n=\sqrt{\frac{p_n}{S_n}\mathrm{PL}(r_n)} h_n x_n + z_n,
%\end{equation}
%where
$x_n$ is the normalized transmit signal and $z_n \sim \mathcal{CN}(0,\sigma^2)$ is the additive white Gaussian noise (AWGN). Based on \eqref{eq:received_signal}, the instantaneous signal-to-noise ratio (SNR) of $U_n$ is expressed as $\gamma_n=\frac{p_n}{S_n\sigma^2}\mathrm{PL}(r_n)|h_n|^2$,
%\begin{equation}
%\gamma_n=\frac{p_n}{S_n\sigma^2}  \mathrm{PL}(r_n)  |h_n|^2,
%\end{equation}
where $|h_n|$ denotes the norm of $h_n$.

%\vspace{-0.3em}

\subsection{Effectiveness Metric Formulation}

QoE is a standard metric to jointly evaluate semantic transmission efficiency and reconstruction quality \cite{QoE}.
In semantic text communications, the QoE for $U_n$ is characterized by semantic similarity $\xi_n$ and rate $\phi_n$, where $\xi_n=\frac{\mathbf{B}(\mathbf{s}_n)\mathbf{B}(\hat{\mathbf{s}}_n)^{\mathsf T}}
{\|\mathbf{B}(\mathbf{s}_n)\|\|\mathbf{B}(\hat{\mathbf{s}}_n)\|}$,
with $\mathbf{B}(\cdot)$ and $\hat{\mathbf{s}}_n$ denoting the BERT sentence embedding and the reconstructed sentence, respectively \cite{DeepSC}, and
$\phi_n=\frac{S_nWI}{d_nL_n}$,
with $I$ denoting the average semantic information \cite{SR}.
% where $\xi_n$ measures the similarity between the transmitted sentence $\mathbf{s}_n$ and the reconstructed sentence $\hat{\mathbf{s}}_n$, and $\phi_n$ represents the delivered semantic information per second \cite{SR}.
% \textcolor{red}{where $\xi_n$ is the BERT-based cosine similarity between the source sentence $\mathbf{s}_n$ and the reconstructed sentence $\hat{\mathbf{s}}_n$ \cite{DeepSC}, and $\phi_n$ is the delivered semantic information per second, determined by \(d_n\) and the allocated symbol rate \(S_nW\) \cite{SR}.}

The conventional QoE model adopts an additive relationship with sigmoid-based satisfaction functions \cite{QoE}, given by $\mathrm{QoE}_n = w_n G_n^{R} + (1-w_n) G_n^{A}$,
%\begin{equation}\label{eq:conventional_QoE}
%\mathrm{QoE}_n = w_n G_n^{R} + (1-w_n) G_n^{A},
%\end{equation}
where $w_n \in (0,1)$ is the user preference weight between semantic rate and similarity, %In \eqref{eq:conventional_QoE}, 
$G_n^{R}\!=\! {1}/{\left(1\!+\!e^{\beta_n(\phi_n^{\mathrm{req}}\!-\!\phi_n)}\right)}$ and $G_n^{A}\!=\!{1}/{\left(1\!+\!e^{\lambda_n(\xi_n^{\mathrm{req}}\!-\!\xi_n)}\right)}$ are satisfaction functions, and $\phi_n^{\mathrm{req}}$ and $\xi_n^{\mathrm{req}}$ denote the required semantic rate and similarity, respectively, while $\beta_n$ and $\lambda_n$ control the steepness of the corresponding sigmoid functions. While providing optimization tractability, this QoE model is inherently compensatory, since high satisfaction in one dimension can artificially offset poor performance in the other.
%Although this QoE model offers optimization tractability, it is compensatory in the sense that high satisfaction in one dimension can partially offset low satisfaction in the other.}

To address this failure, we propose a new QoE formulation that adheres to normalization, monotonicity, a strict zero boundary, and a flexible preference tradeoff. The proposed QoE is defined as
%\begin{equation}\label{eq:qoe_new_inv_orig}
%\mathrm{QoE}_n=\left(1-e^{-\tfrac{1}{w_n}\,\bar{\phi}_n}\right)\left(1-e^{-\tfrac{1}{1-w_n}\,\bar{\xi}_n}\right),
%\end{equation}
\begin{equation}\label{eq:qoe_new_inv_orig}
\mathrm{QoE}_n=\left(1-e^{-\bar{\phi}_n/{w_n}}\right)\left(1-e^{-\bar{\xi}_n/(1-w_n)\,}\right),
\end{equation}
where $\bar{\phi}_n = {\phi_n}/{\phi_n^{\mathrm{req}}}$ and $\bar{\xi}_n = {\xi_n}/{\xi_n^{\mathrm{req}}}$ denote the normalized semantic rate and similarity, respectively. 
This formulation ensures that $\mathrm{QoE}_n=0$ if either $\phi_n=0$ or $\xi_n=0$ and remains non-decreasing in both dimensions, 
% \textcolor{red}{For the proposed QoE, \(w_n\) is restricted to \([\epsilon,1-\epsilon]\), where \(\epsilon>0\) is a small constant,}
while the inverse weights $\frac{1}{w_n}$ and $\frac{1}{1-w_n}$ allow for flexible preference control. Here, the multiplicative formulation is adopted for time-sensitive text semantic services, in which successful service delivery requires both an adequate semantic information delivery rate and sufficiently accurate preservation of the intended meaning, characterized by the semantic rate and similarity, respectively.
%to characterize semantic services that necessitate the simultaneous fulfillment of both semantic rate and similarity. 
Unlike additive QoE, this formulation ensures that a deficiency in either dimension yields a low QoE, effectively preserving the desired non-compensatory behavior. Additionally, $w_n$ allows the framework to prioritize rate or similarity according to specific user needs, thereby capturing heterogeneous semantic requirements across users.

\begin{figure}[!t]
\centering
\subfloat[Conventional QoE ($w=0.2$)]{
\includegraphics[width=0.38\columnwidth]{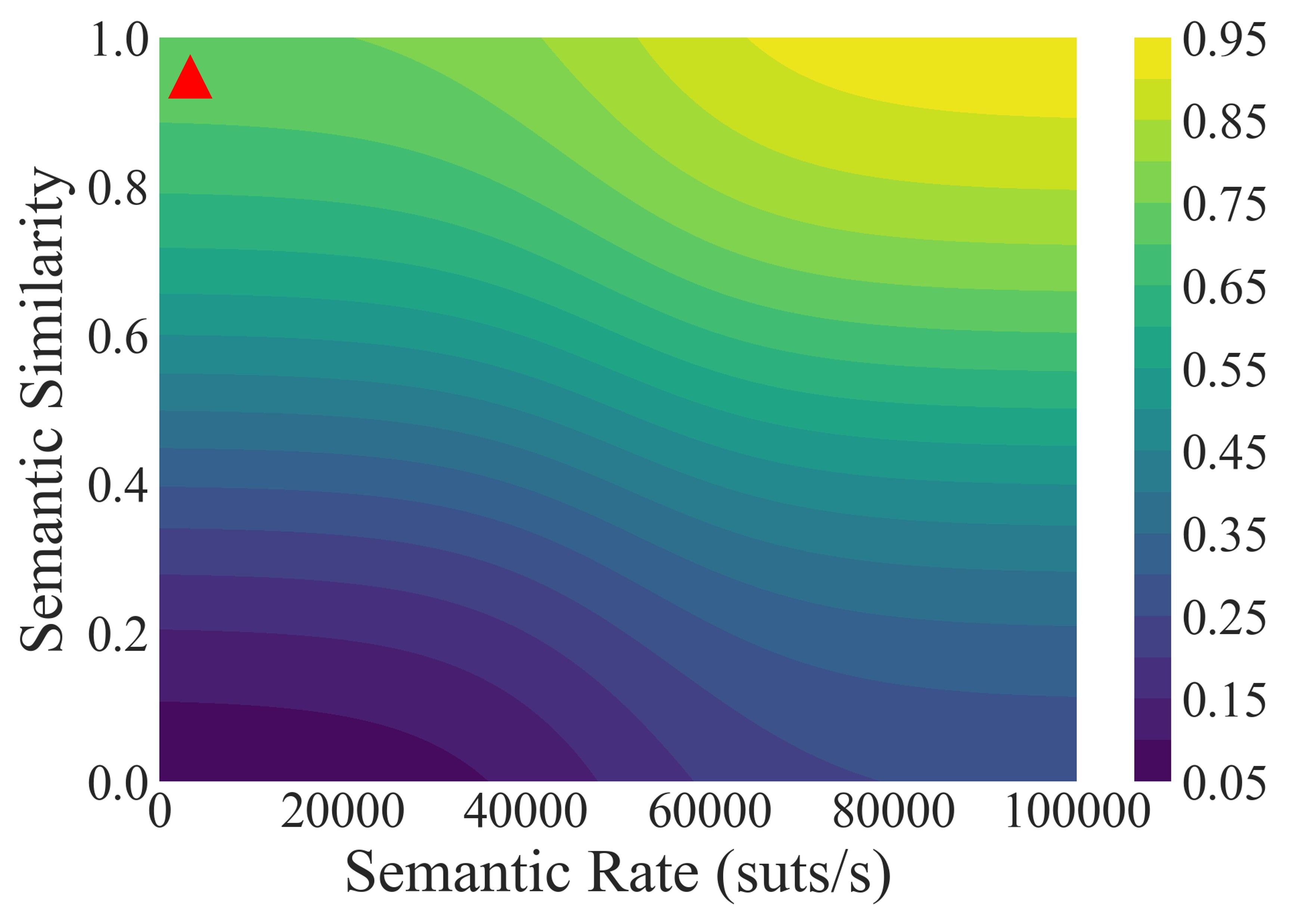}\vspace{-0.5em}
\label{fig:qoe_w02_old}
}\hfill
\subfloat[Proposed QoE ($w=0.2$)]{
\includegraphics[width=0.38\columnwidth]{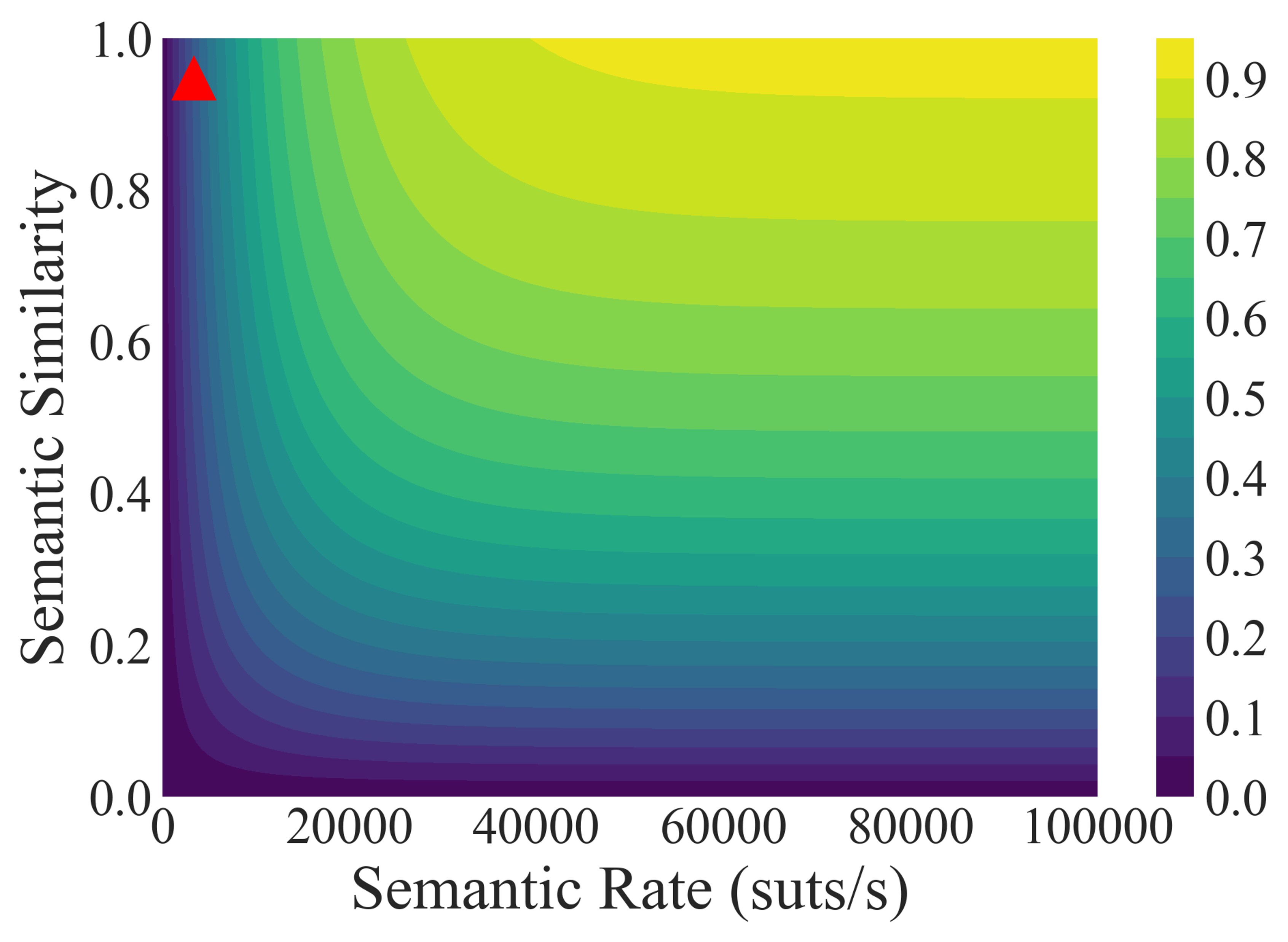}\vspace{-0.5em}
\label{fig:qoe_w02_new}
}\\%[0.3em]
\subfloat[Conventional QoE ($w=0.5$)]{
\includegraphics[width=0.38\columnwidth]{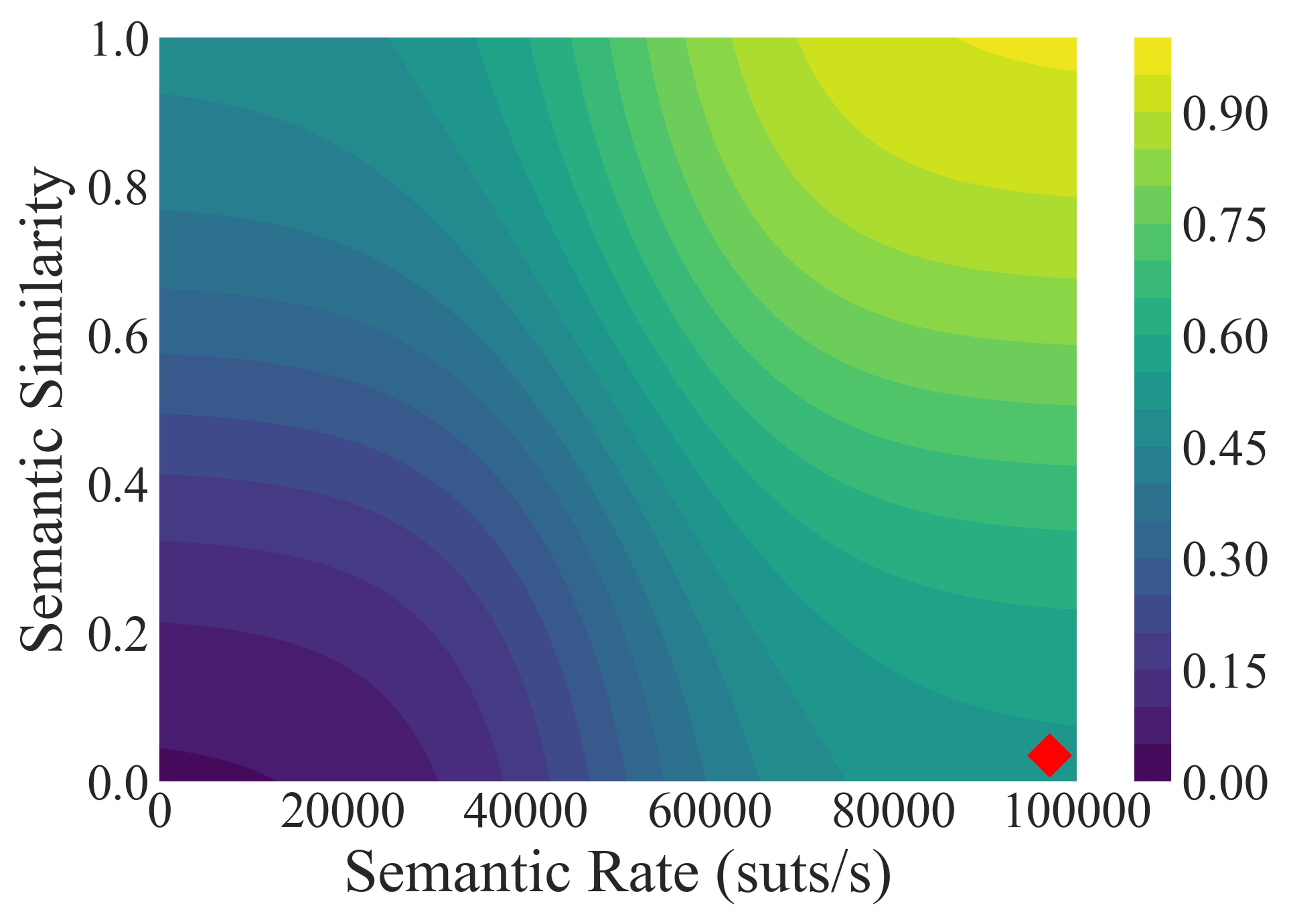}\vspace{-0.5em}
\label{fig:qoe_w05_old}
}\hfill
\subfloat[Proposed QoE ($w=0.5$)]{
\includegraphics[width=0.38\columnwidth]{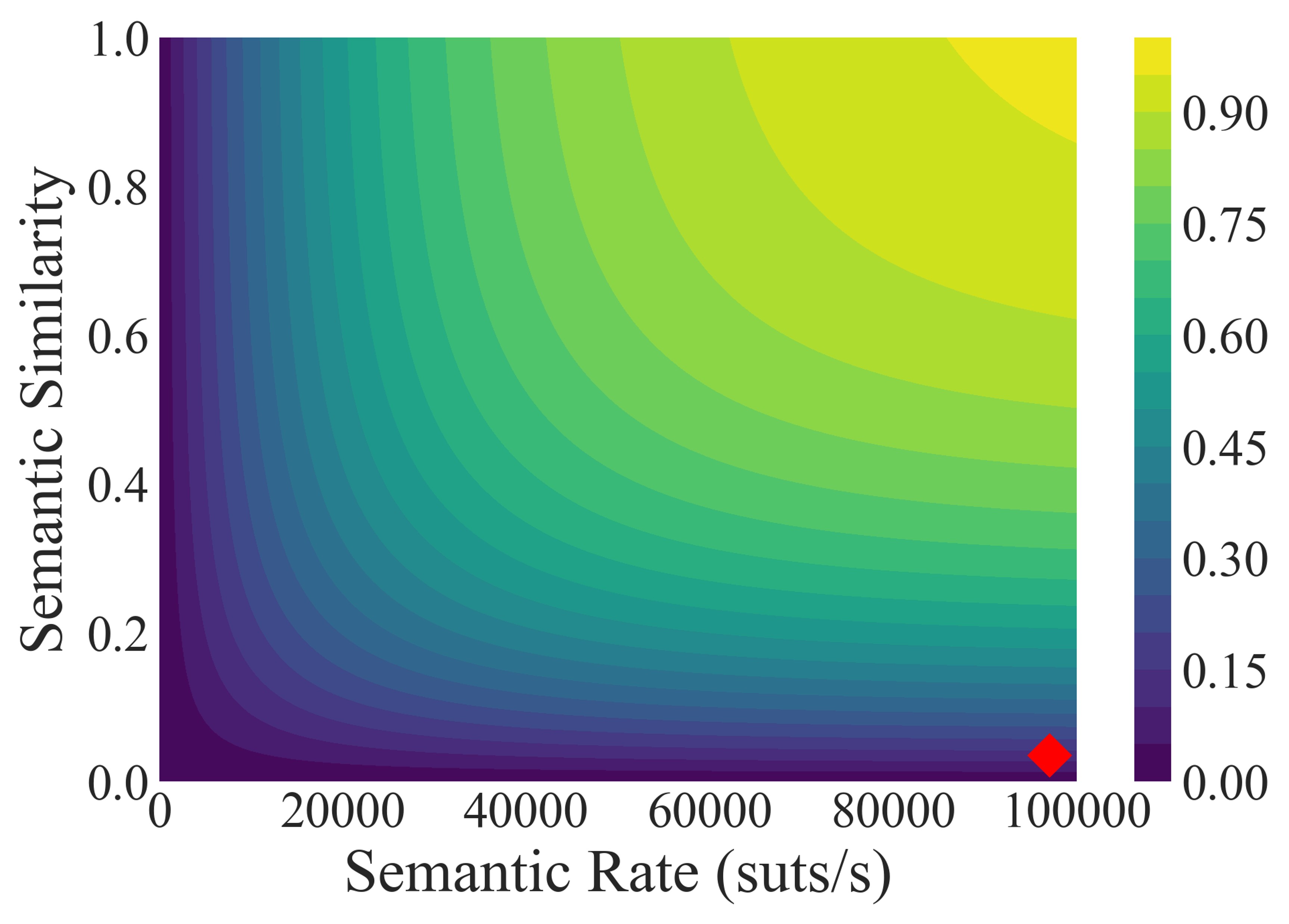}\vspace{-0.5em}
\label{fig:qoe_w05_new}}
\vspace{-0.5em}
\caption{Comparison of conventional QoE and proposed QoE for different preference weights $w_n=w$, $\beta_n=10^{-4}$, and $\lambda_n=5$, $\forall n$, where red triangles and diamonds denote high-similarity low-rate and low-similarity high-rate points, respectively, with high conventional QoE in (a) and (c) and low proposed QoE in (b) and (d).}\vspace{-1.8em}
\label{fig:qoe_heatmaps}
\end{figure}

Fig. \ref{fig:qoe_heatmaps} compares the conventional sigmoid-based QoE with the proposed QoE in \eqref{eq:qoe_new_inv_orig} for $w_n=w=0.2$ and $0.5$, $\forall n$. Due to its additive structure, the conventional model allows for performance compensation, yielding a moderate QoE even when one dimension is critically insufficient. In contrast, the proposed QoE remains low whenever either $\phi$ or $\xi$ is insufficient, effectively enforcing a joint requirement on both dimensions. As $w$ increases, the QoE contours become more symmetric, reflecting a balanced tradeoff. Overall, the proposed QoE provides a bounded and preference-consistent utility that aligns more closely with joint semantic throughput-fidelity requirements.

%\vspace{-0.3em}

\subsection{Optimization Problem Formulation}\label{subsec:opt}

For the considered system, we formulate a resource allocation problem to maximize its sum QoE, subject to coupled resource constraints imposed by shared spectrum and power budgets. The resulting mixed continuous-discrete optimization problem is formulated as
\begin{subequations}\label{eq:P0}
\begin{align}
\mathcal{P}0\quad
\max_{\{p_n,\, S_n,\, d_n\}}
\quad & \sum_{n \in \mathcal{Q}} \mathrm{QoE}_n \\
\mathrm{s.t.}\quad
& p_{\min} \le p_n \le p_{\max}, \quad \forall n \in \mathcal{Q}, \label{eq:C1}\\
& \sum_{n \in \mathcal{Q}} p_n \le P_{\mathrm{tot}}, \label{eq:C2}\\
& 1 \le S_n \le S_{\mathrm{tot}}, \quad \forall n \in \mathcal{Q}, \label{eq:C3}\\
& \sum_{n \in \mathcal{Q}} S_n \le S_{\mathrm{tot}}, \label{eq:C4}\\
& d_n \in \mathcal{D}, \quad \forall n \in \mathcal{Q},\label{eq:C5}
\end{align}
\end{subequations}
where %$p_n \in \mathbb{R}_+$, $S_n \in \mathbb{Z}_+$, and $d_n \in \mathcal{D}$ %denotes the continuous transmit power, integer-valued subcarrier allocation, and discrete semantic configuration, respectively, with
$\mathcal{D}\!=\!\{\!\hat{d}_1,\dots,\hat{d}_M\!\}$ denotes a finite set of $M$ discrete candidates of the average word length %number of semantic symbols per word
for each user.
%generated by DeepSC. 
Constraints \eqref{eq:C1} and \eqref{eq:C2} enforce the per-user and total power limits, \eqref{eq:C3} and \eqref{eq:C4} enforce the per-user and total subcarrier constraints, and \eqref{eq:C5} enforces the discrete semantic decision set.

%the finite DeepSC configuration set.

\section{Reinforcement Learning-Based Solution}

The problem $\mathcal{P}0$ is inherently non-convex due to the coupling among users and hybrid continuous-discrete decision variables, rendering conventional optimization intractable. To address this, 
we develop a reinforcement learning (RL)-based solution leveraging MADDPG \cite{MADDPG}. Although resource allocation is centrally implemented at the BS, each user is modeled as an agent to decompose the joint decision, while the centralized critic captures cross-user coupling.
Each agent outputs an unconstrained hybrid action, specifically the power $\tilde{p}_n$, subcarrier $\tilde{S}_n$, and semantic logits $\tilde{\mathbf{d}}_n$ for $U_n$.

We note that directly applying MADDPG may lead to infeasible actions that violate the coupled resource constraints. To overcome this issue, we propose a differentiable projection layer that maps unconstrained actor outputs to feasible actions, ensuring constraint satisfaction while preserving differentiability for end-to-end training.

%\vspace{-0.5em}

\subsection{Differentiable Projection Layer for Hybrid Resource Allocation}\label{subsec:proj}

We first design projection modules for power, subcarrier, and semantic actions to ensure constraint satisfaction while preserving differentiability, as follows:

\subsubsection{Power projection with total budget and per-user constraint}

For the continuous power action, we enforce the per-user power constraint and the total power budget via a reservation allocation followed by capped redistribution. Specifically,  we compute softmax ratios via $\omega_n={e^{\tilde{p}_n}}/{\sum_{j=1}^{N}e^{\tilde{p}_j}}$, where $\tilde{p}_n$ is produced by the actor network. We then allocate the power as $p_n=p_{\min}+P_{\mathrm{rem}}\omega_n$ with $P_{\mathrm{rem}}= P_{\mathrm{tot}} - N p_{\min}$, to satisfy the minimum and total constraints. To enforce $p_n \le p_{\max}$, users exceeding $p_{\max}$ are capped, and the remaining budget is redistributed over uncapped users via a renormalized softmax. This capped-simplex procedure ensures feasibility and convergency in iterations.

\subsubsection{Subcarrier projection with global sum constraint and minimum allocation}

For the discrete subcarrier action, %similar to the computation of allocated power, the
we compute subcarrier allocation weights via softmax as $\omega_n^{(S)}={e^{\tilde{S}_n}}/{\sum_{j=1}^{N}e^{\tilde{S}_j}}$, yielding the relaxed allocation as $S_n = 1 + (S_{\mathrm{tot}}-N)\omega_n^{(S)}$, which satisfies $S_n \ge 1$ and $\sum_{n=1}^{N}S_n \le S_{\mathrm{tot}}$. For practical execution, the relaxed $S_n$ is rounded to integers with a lightweight repair, where remaining subcarriers are assigned to users with the largest fractional parts and excess allocations are removed from those with the smallest fractional parts, ensuring $S_n \ge 1$ and $\sum_{n=1}^{N}S_n \le S_{\mathrm{tot}}$.

\subsubsection{Discrete semantic symbol-per-word selection via Gumbel-softmax (GS)}

For the discrete semantic action, we find that employing the MADDPG framework \cite{DeepSC}, each agent outputs its $M$ logits, $\tilde{\mathbf{d}}_n=[\tilde{d}_{n,1},\ldots,\tilde{d}_{n,M}]$. %where $\tilde{d}_{n,m}$ denotes the $m$-th logit, $m\in\{1,\ldots,M\}$.
To enable differentiable learning, we adopt the GS relaxation \cite{GS} to compute the selection probabilities as %$\pi_{n,m}=\frac{e^{\frac{\tilde{d}_{n,m}+g_{n,m}}{\tau}}}{\sum_{\ell=1}^{M}e^{\frac{\tilde{d}_{n,\ell}+g_{n,\ell}}{\tau}}}$
%\frac{\exp\left((\tilde{d}_{n,m}+g_{n,m})/\tau\right)}{\sum_{\ell=1}^{M}\exp\left((\tilde{d}_{i,\ell}+g_{i,\ell})/\tau\right)},
$\pi_{n,m}=e^{\frac{\tilde{d}_{n,m}+g_{n,m}}{\tau}}/\big({\sum_{\ell=1}^{M}e^{\frac{\tilde{d}_{n,\ell}+g_{n,\ell}}{\tau}}}\big)$, where $g_{n,m}$ is Gumbel noise with $m\in\{1,\ldots,M\}$, and $\tau$ is the temperature. The relaxed output is given by $d_n^{\mathrm{soft}}=\sum_{m}\pi_{n,m}\hat{d}_m$ for backpropagation. A hard decision of $d_n=\hat{d}_{m_i^{\ast}}$ with $m_i^{\ast}=\arg\max_m \pi_{n,m}$ is made for execution. %We emphasize that all evaluation results are obtained using repaired subcarrier allocations and the hard decision, explained in the aforementioned modules, thereby accounting for the relaxation-to-execution effect and satisfying constraints \eqref{eq:C3}--\eqref{eq:C5}.

%\vspace{-0.5em}

\subsection{Adaptive Performance-Driven Temperature Annealing}\label{subsec:apta}

The temperature $\tau$ in GS relaxation governs the exploration-exploitation tradeoff in discrete semantic selection. Conventional schemes typically adopt fixed exponential decay schedules, often failing to adapt to learning dynamics. To overcome this, we propose an APTA scheme to dynamically adjust $\tau$ based on QoE evolution. This avoids the premature discretization caused by rapid decay and the excessive exploration and delayed convergence resulting from slow decay.

%In particular, rapid decay leads to premature discretization, while slow decay causes excessive exploration and delayed convergence. To overcome this, we propose an APTA scheme, which dynamically adjusts $\tau$ based on QoE evolution.

We denote $\mathrm{QoE}_{t}$ as the instantaneous QoE at iteration $t$ and define its exponential moving average as $
Z_{t} = (1-\rho_{\mathrm{ema}})Z_{t-1} + \rho_{\mathrm{ema}}\mathrm{QoE}_{t}
$, where $\rho_{\mathrm{ema}}\in(0,1)$ is the smoothing factor. Here, the exponential moving average is used to smooth the instantaneous QoE trajectory, ensuring that temperature adaptation tracks the underlying training trend rather than short-term stochastic fluctuations. Based on $Z_{t}$, we define the deficit and improvement indicators as $\Lambda^{\mathrm{def}}_{t}=\max\left(0,\,Q_{\mathrm{target}}-Z_{t}\right)$ and $\Lambda^{\mathrm{imp}}_{t}=\max\left(0,\,Z_{t}-Q_{\min}\right)$, respectively, where $Q_{\mathrm{target}}$ and $Q_{\min}$ denote the target and minimum QoE thresholds, respectively. The adaptive decay factor at iteration $t$ is then computed as
\begin{equation}\label{eq:rho_t}
\rho_{t}=\rho_{\min}+(\rho_{\max}-\rho_{\min}) \left(1-e^{-\kappa_{\mathrm{def}}\Lambda^{\mathrm{def}}_{t}}\right) e^{-\kappa_{\mathrm{imp}}\Lambda^{\mathrm{imp}}_{t}},
\end{equation}
where $\rho_{\min}$ and $\rho_{\max}$ are the minimum and maximum decay rates, and $\kappa_{\mathrm{def}}$ and $\kappa_{\mathrm{imp}}$ control the sensitivity to the deficit and improvement indicators. Jointly considering \eqref{eq:rho_t} and the temperature at iteration $t$, denoted by $\tau_{t}$, the temperature at iteration $t+1$ is updated as $
\tau_{t+1} = \max\left(\tau_{\min}, \rho_{t}\tau_{t}\right)
$, where $\tau_{\min}$ is the minimum allowable temperature. Thus, a larger $\Lambda^{\mathrm{def}}_{t}$ increases $\rho_{t}$ toward $\rho_{\max}$ to retard temperature decay and preserve exploration, whereas a larger $\Lambda^{\mathrm{imp}}_{t}$ drives $\rho_{t}$ toward $\rho_{\min}$, accelerating the temperature decrease and enabling the policy to commit more rapidly to discrete decisions.

We next provide a theoretical convergence analysis of the GS-based policy optimization to characterize the impact of a general temperature schedule, denoted by $\{\tau_t\}_{t=0}^{T-1}$, on training dynamics. To facilitate this analysis, we first formalize the discrete and relaxed objectives. During inference, each agent executes a discrete semantic action. We define $J(\theta)\triangleq \mathbb{E}[F(\mathbf{a}(\theta))]$ as the expected return, where $F(\mathbf{a}(\theta))$ is the reward associated with the discretized joint action  $\mathbf{a}(\theta)$ produced by the policy parameterized by $\theta$, with $\theta$ being the collection of all trainable actor parameters. During training, the discrete actions are replaced by the GS relaxation with temperature $\tau>0$. This yields a differentiable joint action $\mathbf{a}_\tau(\theta)$ and the corresponding relaxed objective $J_\tau(\theta)\triangleq \mathbb{E}[F(\mathbf{a}_\tau(\theta))]$.

In our analysis, we
%\textit{Assumptions:} We
consider stochastic gradient updates $\theta_{t+1}=\theta_t+\eta g_t$ with $\eta \leq 1/\mathcal{L}$, where $\eta$ is the learning rate, $g_t$ is the stochastic gradient estimator at iteration $t$, and %$\mathcal{L}$ is the smoothness constant of the relaxed objective}.
the relaxed objective $J_{\tau}(\theta)$ is assumed to be $\mathcal{L}$-smooth. Also, we consider the stochastic gradients to be unbiased with bounded variance, satisfying $\mathbb{E}[g_t|\theta_t]=\nabla J_{\tau_t}(\theta_t)$ and $\mathbb{E}[\|g_t-\nabla J_{\tau_t}(\theta_t)\|^2|\theta_t]\le \sigma^2(\tau_t)$, where $\sigma^2(\tau)$ is the variance bound of the stochastic gradient estimator at temperature $\tau$, which is nonincreasing in $\tau$. Moreover, the relaxation bias satisfies $|J(\theta)-J_\tau(\theta)|\le \varepsilon(\tau)$, where $\varepsilon(\tau)$ is the squared relaxation bias caused by replacing the original objective $J(\theta)$ with the relaxed objective $J_{\tau}(\theta)$, which vanishes as $\tau\rightarrow0$. Furthermore, the gradient transfer satisfies $\|\nabla J(\theta)\|\le \|\nabla J_\tau(\theta)\| + c\,\varepsilon(\tau)$, where the positive constant $c$ quantifies the impact of the relaxation bias on gradient mismatch. Finally, the objective drift due to temperature variation is bounded as $|J_\tau(\theta)-J_{\tau'}(\theta)|\le L_\tau|\tau-\tau'|$ \cite{Maddison2017}, where $L_\tau$ is the Lipschitz constant of the objective with respect to $\tau$ and $\tau'$ is another temperature used for comparison. Under these considerations, we derive the %schedule-dependent
convergence bound in the following proposition:

\begin{Proposition}\label{Proposition:1}%[Schedule-Dependent Convergence Bound]
%Under the above assumptions, for
For a given temperature schedule $\{\tau_t\}_{t=0}^{T-1}$, the trainable actor parameters at iteration $t$, $\theta_t$, must satisfy
\begin{equation}\label{eq:Proposition1_S}
\frac{1}{T}\sum_{t=0}^{T-1}\mathbb{E}\|\nabla J(\theta_t)\|^2\le\frac{4\Delta_0}{\eta T}+2\varpi(\{\tau_t\}),
\end{equation}
where $\Delta_0$ is the initial optimality gap and $\varpi(\{\tau_t\})$ is the schedule-dependent cost, given by
%\begin{align}\label{eq:Proposition1_O}
%\varpi(\{\tau_t\})=&\frac{1}{T}\sum_{t=0}^{T-1}\sigma^2(\tau_t)+
%\frac{c^2}{T}\sum_{t=0}^{T-1}\varepsilon^2(\tau_t)\notag\\
%&+\frac{2L_\tau}{\eta T}\sum_{t=0}^{T-1}\left|\tau_{t+1}-\tau_t\right|.
%\end{align}
% \begin{align}\label{eq:Proposition1_O}
% \varpi(\{\tau_t\})\!=\!\frac{1}{T}\sum_{t=0}^{T-1}\left(\sigma^2(\tau_t)\!+\!
% c^2\varepsilon^2(\tau_t)\right)\!+\!\frac{2L_\tau}{\eta T}\sum_{t=0}^{T-1}\left|\tau_{t\!+\!1}\!-\!\tau_t\right|.
% \end{align}
\begin{align}\label{eq:Proposition1_O}
\varpi(\{\tau_t\})\!=\!\frac{1}{T}\sum_{t=0}^{T-1}\left(\sigma^2(\tau_t)\!+\!
c^2\varepsilon^2(\tau_t)\!+\!\frac{2L_\tau}{\eta}\left|\tau_{t\!+\!1}\!-\!\tau_t\right|\right).
\end{align}
\begin{IEEEproof}
Since $J_{\tau_t}(\theta)$ is $\mathcal{L}$-smooth, applying \cite[Lemma 3.2]{Lsmooth} to the stochastic update, we obtain
%\begin{align}\label{eq:ineq1}
%\mathbb{E}\left[J_{\tau_t}(\theta_{t+1})|\theta_t\right]
%\ge&J_{\tau_t}(\theta_t)+\left(\eta-\frac{\mathcal{L}}{2}\eta^2\right) \left\|\nabla J_{\tau_t}(\theta_t)\right\|^2\notag\\
%&-\frac{\mathcal{L}}{2}\eta^2 \sigma^2(\tau_t),
%\end{align}
\begin{align}\label{eq:ineq1}
\mathbb{E}\!\left[\!J_{\tau_t}\!(\!\theta_{t\!+\!1}\!)|\theta_t\!\right]
\!\!\ge\!\! J_{\tau_t}\!(\!\theta_t\!)\!+\!\left(\!\!\eta\!-\!\frac{\mathcal{L}}{2}\eta^2\!\!\right)\! \!\left\|\nabla J_{\tau_t}(\!\theta_t\!)\!\right\|^2\!-\!\frac{\mathcal{L}}{2}\!\eta^2\! \sigma^2(\!\tau_t\!),
\end{align}
where we invoke the unbiasedness and bounded-variance properties of stochastic gradients. Given $\eta \le 1/\mathcal{L}$, \eqref{eq:ineq1} becomes
\begin{align}\label{eq:ineq2}
\frac{\eta}{2}\! \mathbb{E}\left[\!\|\nabla J_{\tau_t}(\theta_t)\|^2\!\right]\!\le\!\mathbb{E}\left[\!J_{\tau_t}(\theta_{t\!+\!1})\!\right]\! -\!\mathbb{E}\left[J_{\tau_t}(\theta_t)\right]\!+\!\frac{\eta}{2}\sigma^2(\tau_t).
\end{align}
To account for the time-varying temperature, we observe that the objective drift between iterations satisfies
\begin{align}\label{eq:ineq3}
\left|J_{\tau_{t+1}}(\theta_{t+1})-J_{\tau_t}(\theta_{t+1})\right|\le
L_\tau\left|\tau_{t+1}-\tau_t\right|.
\end{align}
Summing \eqref{eq:ineq2} from $t=0$ to $T-1$, telescoping the resulting sequence, and incorporating \eqref{eq:ineq3}, we obtain %the drift term in
%we define the potential sequence as $\Phi_t \triangleq J_{\tau_t}(\theta_t)$. By adding and subtracting $J_{\tau_{t+1}}(\theta_{t+1})$ and leveraging the drift bound, we obtain
%\begin{align}\label{eq:ineq3}
%\left|J_{\tau_{t+1}}(\theta_{t+1}) - J_{\tau_t}(\theta_{t+1})\right|
%\le
%L_\tau \left|\tau_{t+1} - \tau_t\right|.
%\end{align}
%Calculating the summation of \eqref{eq:ineq2} from $t=0$ to $T-1$ and telescoping the resulting sequence, accounting for the drift term in \eqref{eq:ineq3}, we obtain
\begin{align}\label{eq:ineq4}
\frac{1}{T}\sum_{t=0}^{T-1}\mathbb{E}\|\nabla J_{\tau_t}(\theta_t)\|^2\le&
\frac{2\Delta_0}{\eta T}+
\frac{1}{T}\sum_{t=0}^{T-1}\sigma^2(\tau_t)\notag\\
&+\frac{2L_\tau}{\eta T}\sum_{t=0}^{T-1}\left|\tau_{t+1}-\tau_t\right|.
\end{align}
Finally, applying the gradient transfer and relaxation bias bounds, we obtain
\begin{align}\label{eq:ineq5}
\|\nabla J(\theta_t)\|^2
\le 2\|\nabla J_{\tau_t}(\theta_t)\|^2
+2c^2 \varepsilon^2(\tau_t).
\end{align}
Averaging \eqref{eq:ineq5} over $T$ and combining the averaged result with \eqref{eq:ineq4} yields \eqref{eq:Proposition1_S}.%See Appendix~\ref{Appendix:A}.
\end{IEEEproof}
\end{Proposition}

\emph{Proposition \ref{Proposition:1}} demonstrates that the convergence is governed by $\varpi(\{\tau_t\})$, which captures variance accumulation, relaxation bias, and schedule-induced temperature drift. Hence, faster convergence can be achieved by reducing these three components. To establish the superiority of the proposed APTA scheme, it suffices to show that it yields a smaller $\varpi(\{\tau_t\})$ than the conventional scheme with fixed decay schedules, as proven in the following theorem.

% \vspace*{0.5mm}

\begin{Theorem}\label{Proposition:2}
We assume that $\sigma^2(\tau)=a\tau^{-p}$ and $\varepsilon^2(\tau)=b\tau^{q}$ with $a$, $b$, $p$, $q>0$. Under the APTA scheme, the temperature schedule is adaptively adjusted across two phases, formulated as
\begin{align}\label{eq:APTA_temp}
\tau_t^{\mathrm{APTA}}=
\begin{cases}
\chi_\psi\tau_t^{\mathrm{fix}}, & \textrm{if }t\in \psi,\\
\chi_{\bar{\psi}}\tau_t^{\mathrm{fix}}, & \textrm{otherwise},
\end{cases}
\end{align}
where $\tau_t^{\mathrm{fix}}$ is the temperature at iteration $t$ for the conventional scheme. In \eqref{eq:APTA_temp}, we denote $\psi=\{0,\ldots,t^\star-1\}$ as the early phase and $\bar{\psi}=\{t^\star,\ldots,T\}$ as the late phase, where $t^\star$ is the switching iteration index, and further denote $\chi_\psi=1+\delta_\psi$ and $\chi_{\bar{\psi}}=1-\delta_{\bar{\psi}}$ as the temperature scaling factors in the early and late phases, respectively, with $\delta_\psi>0$ and $\delta_{\bar{\psi}}>0$ characterizing adjustment magnitudes. We clarify that a sufficient condition to satisfy $\varpi(\{\tau_t^{\mathrm{APTA}}\}) \le \varpi(\{\tau_t^{\mathrm{fix}}\})$ is
\begin{equation}\label{eq:apta_condition_main}
a p(\delta_{\bar{\psi}} V_{\bar{\psi}}-\delta_\psi V_\psi)+c^2 b q(\delta_\psi H_\psi-\delta_{\bar{\psi}} H_{\bar{\psi}})\le 0,
\end{equation}
where $V_\psi=\sum_{t\in \psi}(\tau_t^{\mathrm{fix}})^{-p}, V_{\bar{\psi}}=\sum_{t\in \bar{\psi}}(\tau_t^{\mathrm{fix}})^{-p}, H_\psi=\sum_{t\in \psi}(\tau_t^{\mathrm{fix}})^{q}$, and $H_{\bar{\psi}}=\sum_{t\in \bar{\psi}}(\tau_t^{\mathrm{fix}})^{q}$ denote the corresponding phase-wise moments.
\begin{IEEEproof}
We compare the APTA scheme and the conventional scheme through the schedule-dependent cost $\varpi(\{\tau_t\})$. To isolate the variance-bias tradeoff, we define the drift-free component $\varpi_0(\cdot)$ by omitting the temperature drift term, i.e. the last term in \eqref{eq:Proposition1_O}. Given $\sigma^2(\tau)=a\tau^{-p}$ and $\varepsilon^2(\tau)=b\tau^{q}$, the difference $\Delta\varpi_{0}\triangleq\varpi_0^{\mathrm{APTA}}(\{\tau_t\})-\varpi_0^{\mathrm{fix}}(\{\tau_t\})$ is given by
\begin{align}\label{eq:prop2_1}
\Delta\varpi_{0}=&\frac{1}{T}\left[a\left(\chi_\psi^{-p}-1\right)V_\psi+ a\left(\chi_{\bar{\psi}}^{-p}-1\right)V_{\bar{\psi}}\right.\notag\\
&\quad\left.+c^2b\left(\chi_\psi^{q}-1\right)H_\psi+ c^2b\left(\chi_{\bar{\psi}}^{q}-1\right)H_{\bar{\psi}}\right].
%\varpi_0^{\mathrm{apta}}(\{\tau_t\})-\varpi_0^{\mathrm{fix}}(\{\tau_t\})\notag\\
%&=\frac{1}{T}\big(
%a(\chi_\psi^{-p}-1)\sum_{t\in \psi}(\tau_t^{\mathrm{fix}})^{-p}+a(\chi_{\bar{\psi}}^{-p}-1)\sum_{t\notin \psi}(\tau_t^{\mathrm{fix}})^{-p}
%\notag\\
%&\quad+c^2b(\chi_\psi^{q}-1)\sum_{t\in \psi}(\tau_t^{\mathrm{fix}})^{q}+c^2b(\chi_{\bar{\psi}}^{q}-1)\sum_{t\notin \psi}(\tau_t^{\mathrm{fix}})^{q} \big).
\end{align}
When $\delta_\psi$ and $\delta_{\bar{\psi}}$ are small, %For small perturbations $\chi_\psi=1+\delta_\psi$ and $\chi_{\bar{\psi}}=1-\delta_{\bar{\psi}}$,
the first-order approximation of \eqref{eq:prop2_1} is obtained as
\begin{align}
\Delta\varpi_{0}
%\varpi_0^{\mathrm{apta}}(\{\tau_t\})-\varpi_0^{\mathrm{fix}}(\{\tau_t\})
\!\approx\!&\frac{1}{T}\left[\!ap\left(\delta_{\bar{\psi}}V_{\bar{\psi}}\!-\!\delta_\psi V_\psi\!\right)\!+\!c^2 b q\left(\delta_\psi H_\psi\!-\!\delta_{\bar{\psi}}H_{\bar{\psi}}\right)\right],
\end{align}
which establishes the sufficient condition in \eqref{eq:apta_condition_main}. Finally, since the temperature drift remains controlled under bounded multiplicative updates, we prove that the APTA scheme achieves a smaller $\varpi(\{\tau_t\})$ than the conventional scheme.
%, completing the proof.
\end{IEEEproof}
\end{Theorem}

\emph{Theorem \ref{Proposition:2}} demonstrates that the APTA scheme reduces the schedule-dependent cost $\varpi(\{\tau_t\})$ by balancing the variance-bias tradeoff across training phases, where a larger temperature suppresses variance in the early phase, while a smaller temperature mitigates relaxation bias in the later phase. Since the drift term is controlled under bounded updates, \eqref{eq:apta_condition_main} provides a practical criterion for performance improvement. Together with \emph{Proposition \ref{Proposition:1}}, we confirm that the APTA scheme achieves a lower stationarity bound and faster convergence than the conventional scheme with fixed decay schedules.

\section{Simulation Results}

In this section, we present simulation results to evaluate the proposed multi-agent hybrid resource allocation framework for downlink OFDMA semantic communication systems. We first assess the effectiveness of the QoE-oriented design, projection-based feasibility handling, and the APTA scheme via QoE learning curves. We then investigate the impact of the number of users on the convergence behavior and QoE performance. For such investigations, four action-handling strategies are considered: (i) Heuristic, which equally allocates resources with uniform sampling of $d_n$, (ii) Penalty+GS (fixed $\tau$), which enforces constraints using penalty terms, (iii) Proj+GS (fixed $\tau$), which applies projection with a fixed GS temperature, and (iv) Proj+GS (APTA), which adopts the proposed adaptive temperature scheme. The system parameters used in simulations are summarized in Table \ref{tab:sim_params_globecom}, unless otherwise specified.

\begin{table}[!t]
\vspace{0.4cm}
\centering
\caption{Key Simulation and Training Parameters}
\vspace{-0.8em}
\label{tab:sim_params_globecom}
\setlength{\tabcolsep}{2pt}
\renewcommand{\arraystretch}{0.9}
\scriptsize
\begin{tabular}{|l|l|l|}
\hline
\textbf{Parameters} & \textbf{Symbols} & \textbf{Values} \\ \hline
Number of users and subcarriers & $N$, $S_{\mathrm{tot}}$ & $3$, $64$ \\ \hline
Subcarrier bandwidth & $W$ & $15$ kHz \\ \hline
Power budget and bounds & $P_{\mathrm{tot}}$, $p_{\min}$, $p_{\max}$ & $3$ mW, $0.1$ mW, $1.5$ mW \\ \hline
Path loss exponent & $\alpha$ & $2.5$ \\ \hline
User distance range & $r_{\textrm{min}}$, $r_{\textrm{max}}$ & $660$ m, $1600$ m \\ \hline
Reference distance & $r_0$ & $1$ m \\ \hline
%Path loss at the reference distance & $\mathrm{PL}(r_0)$ & $7.04 \times 10^{-4}$
%\\ \hline
Carrier frequency  &  $f_c$  &  $900$ MHz
\\ \hline
Noise power & $\sigma^2$ & $-132$ dBm\\ \hline
Symbol-per-word set size & $|\mathcal{D}|$ & $15$ \\ \hline
GS initial / minimum temperature & $\tau_0$, $\tau_{\min}$ & $2$, $0.1$ \\ \hline
Required semantic rate & $\phi^{\mathrm{req}}$ & $20000$ suts/s \\ \hline
Required similarity & $\xi^{\mathrm{req}}$ & $0.75$ \\ \hline
QoE thresholds & $Q_{\mathrm{target}}$, $Q_{\min}$ & $0.9$, $0.7$ \\ \hline
Sensitivity factors & $\kappa_{\mathrm{def}}$, $\kappa_{\mathrm{imp}}$ & $2$, $2$ \\ \hline
Decay factors & $\rho_{\min}$, $\rho_{\max}$, $\rho_{\mathrm{fixed}}$ & $0.8$, $0.99$, $0.9$ \\ \hline
%Total training timesteps & $T$ & $3000$
%\\ \hline
%Learning rate and mini-batch size  &  $\eta,$ -- &
%$3\times10^{-4},256$ \\ \hline
%Replay-buffer capacity & -- &  $10^5$  \\ \hline
%Learning-start timesteps & -- &  $1000$ \\ \hline

\end{tabular}
\vspace{-3em}
\end{table}

%Figs. \ref{fig:qoe_train_eval_curves}(a) and \ref{fig:qoe_train_eval_curves}(b) plot the training and evaluation QoE, respectively, versus timestep for different action-handling strategies. We first observe that under both the proposed and conventional QoE designs, the projection-based strategies converge to stable, high-performance regimes, while the Heuristic and Penalty+GS (fixed $\tau$) strategies consistently exhibit inferior performance. This is because projection-based strategies guarantee feasibility by construction, leading to more stable convergence and improved performance. Moreover, projection-based strategies exhibit significantly improved stability compared with penalty-based strategies. Specifically, Penalty+GS (fixed $\tau$) shows noticeable oscillations and abrupt performance degradation, while Proj+GS (fixed $\tau$) and Proj+GS (APTA) converge smoothly to stable plateaus due to their feasibility-preserving nature. Among the projection-based strategies, Proj+GS (APTA) achieves the fastest convergence, which validates the theoretical results in \emph{Theorem \ref{Proposition:2}}, as APTA balances exploration and discretization during training.

\begin{figure}[!t]
\centering
\begin{subfigure}[t]{\columnwidth}
\centering
\includegraphics[width=0.7\linewidth]{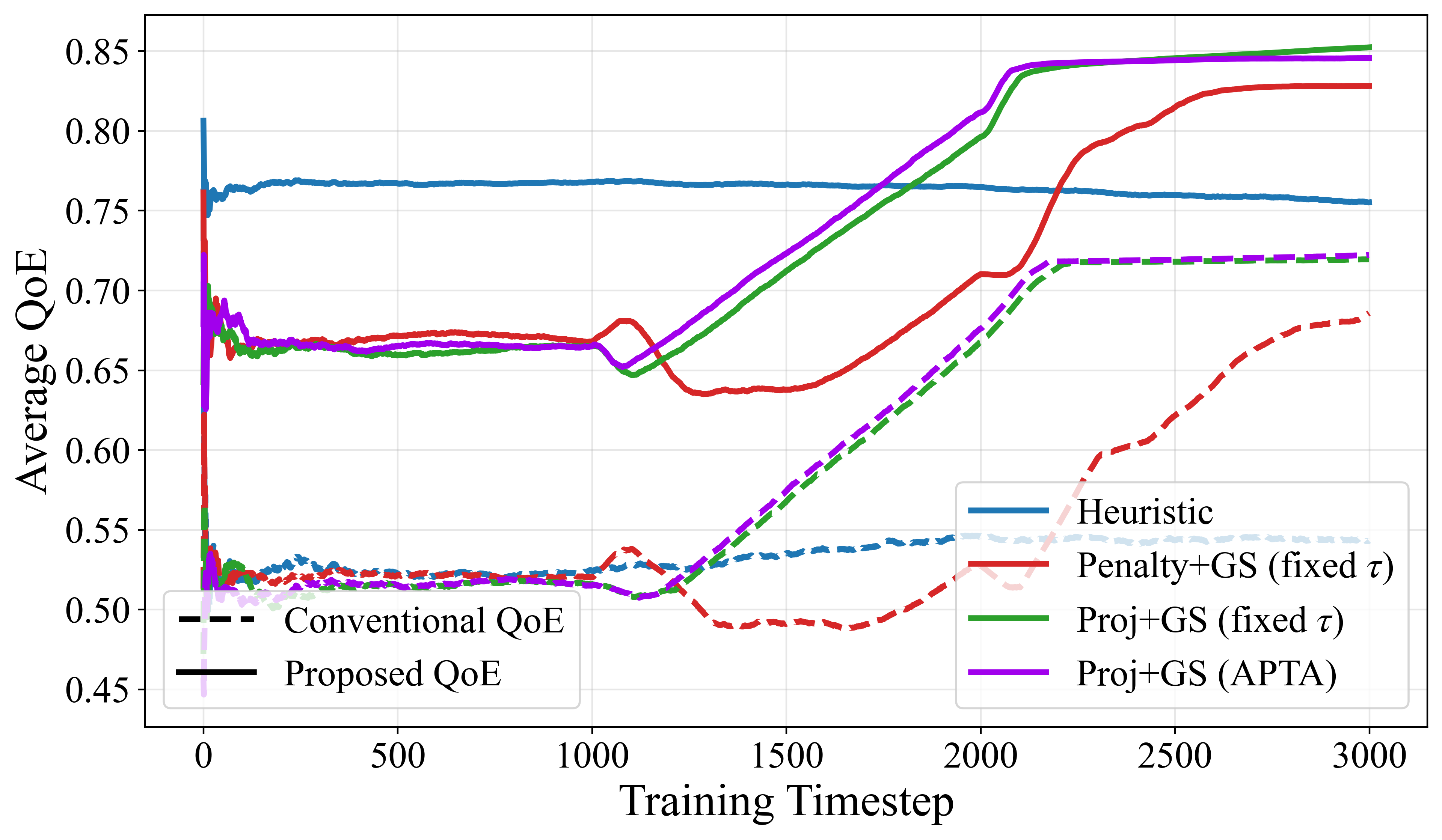}\vspace{-0.5em}
\caption{Training learning curves.}
\label{fig:qoe_train_curves}
\end{subfigure}
%\vspace{0.6em}
\begin{subfigure}[t]{\columnwidth}
\centering
\includegraphics[width=0.7\linewidth]{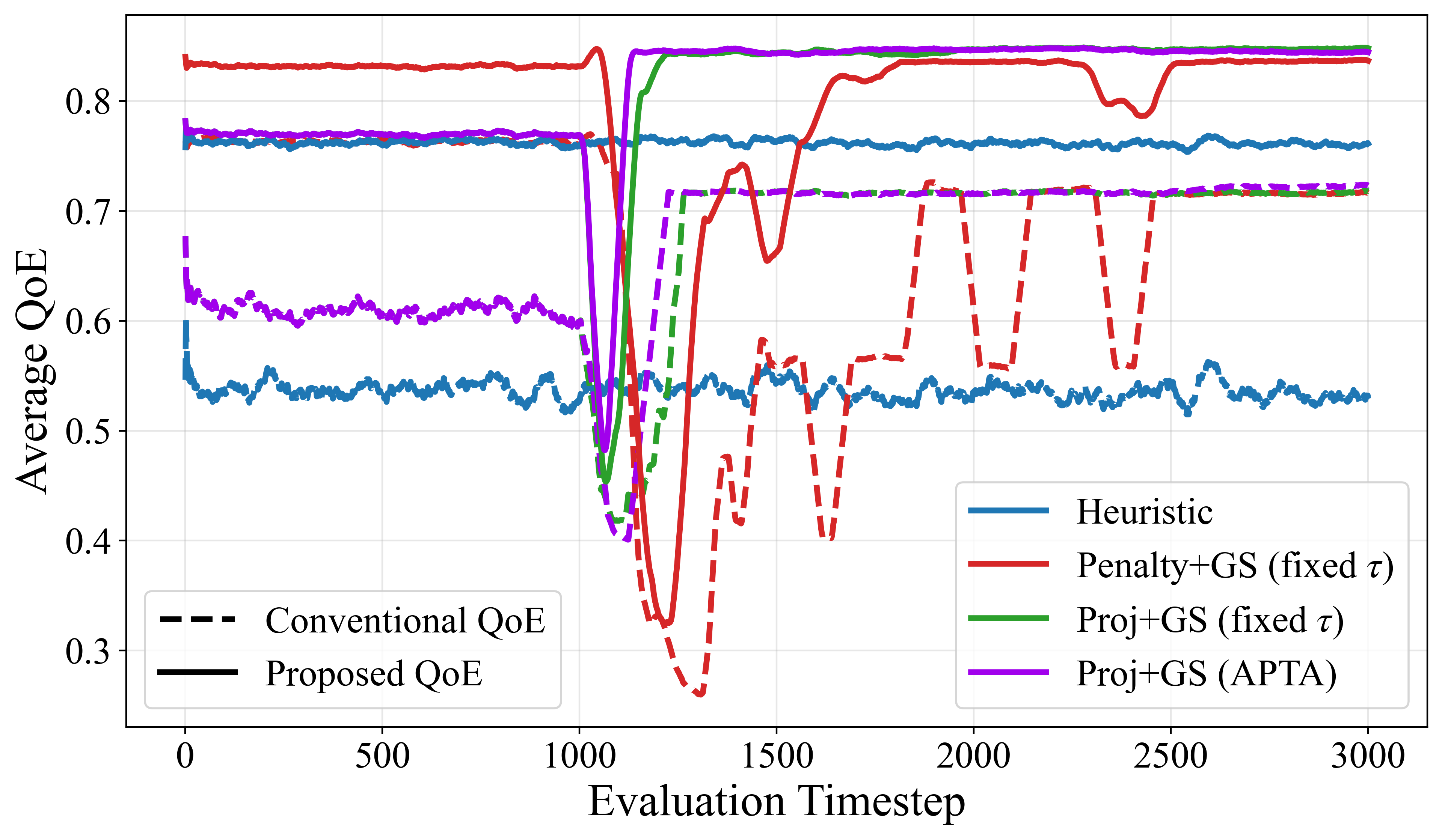}\vspace{-0.5em}
\caption{Evaluation learning curves.}
\label{fig:qoe_eval_curves}
\end{subfigure}\vspace{-0.5em}
\caption{Training and evaluation learning curves of average QoE versus evaluation timestep for different action-handling strategies.}\vspace{-1em}
\label{fig:qoe_train_eval_curves}
\end{figure}

Figs. \ref{fig:qoe_train_eval_curves}(a) and \ref{fig:qoe_train_eval_curves}(b) plot the training and evaluation QoE, respectively, for different action-handling strategies. We first observe that under both the proposed and conventional QoE designs, the projection-based strategies converge to stable, high-performance regimes, whereas the Heuristic and Penalty+GS (fixed $\tau$) strategies achieve inferior performance. By directly mapping the actor outputs onto the feasible resource-allocation region, the projection-based strategy effectively mitigates performance fluctuations and abrupt QoE degradation rather than relying on reward penalties to learn constraint satisfaction. Moreover, we observe that Proj+GS (APTA) converges faster than Proj+GS (fixed ($\tau$)), because APTA promotes sufficient exploration during the early training stage and gradually produces sharper semantic decisions as training progresses, aligning with \emph{Theorem \ref{Proposition:2}}. Overall, the projection layer mainly improves feasibility and training stability, while APTA mainly accelerates the learning of discrete semantic decisions.

\begin{figure}[!t]
\centering
\begin{subfigure}[t]{\columnwidth}
\centering
\includegraphics[width=0.7\linewidth]{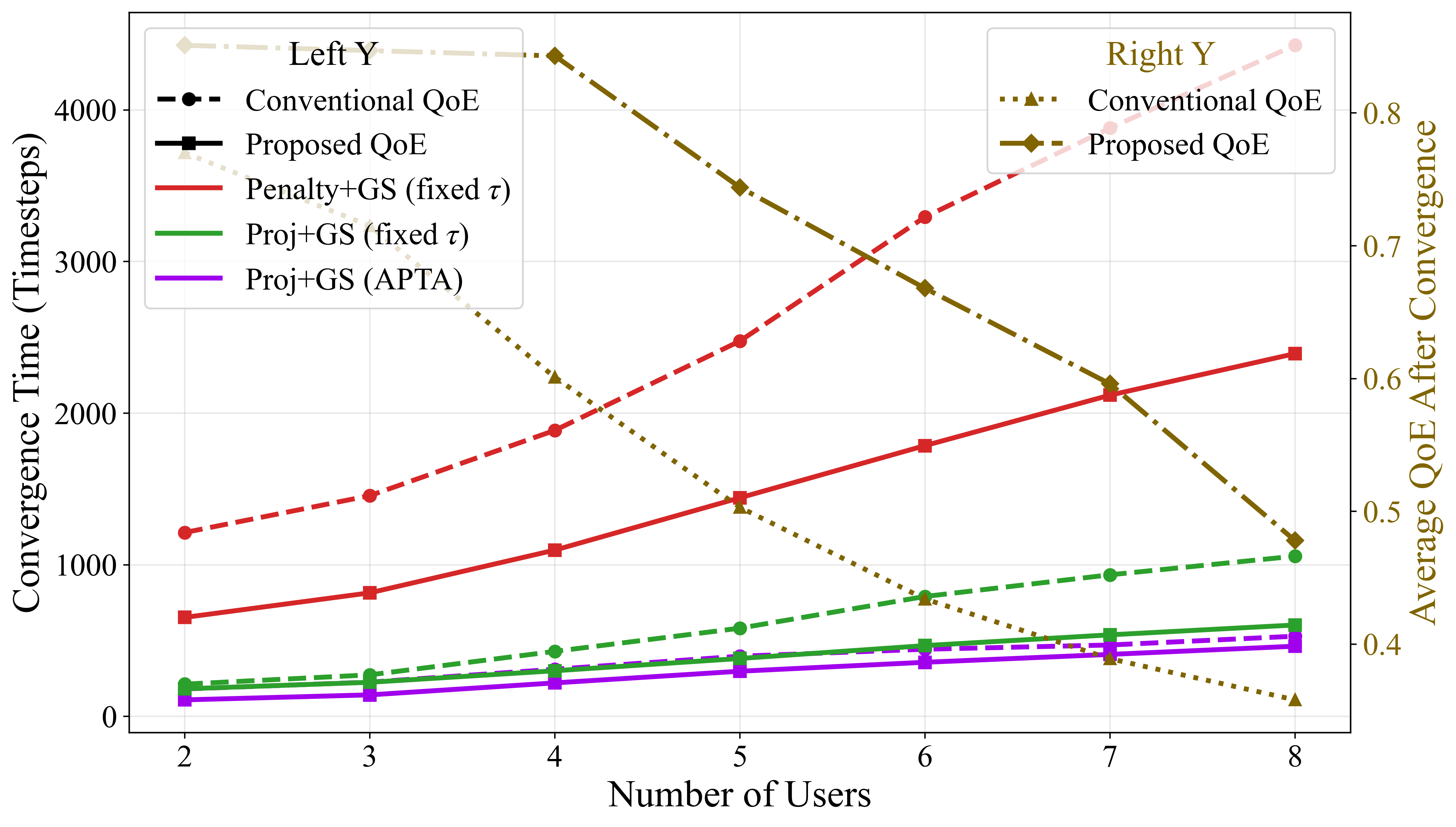}\vspace{-0.5em}
\caption{Curves for preference weight $w_n=0.5$.}
\label{fig:qoe_train_curves}
\end{subfigure}
%\vspace{0.6em}
\begin{subfigure}[t]{\columnwidth}
\centering
\includegraphics[width=0.7\linewidth]{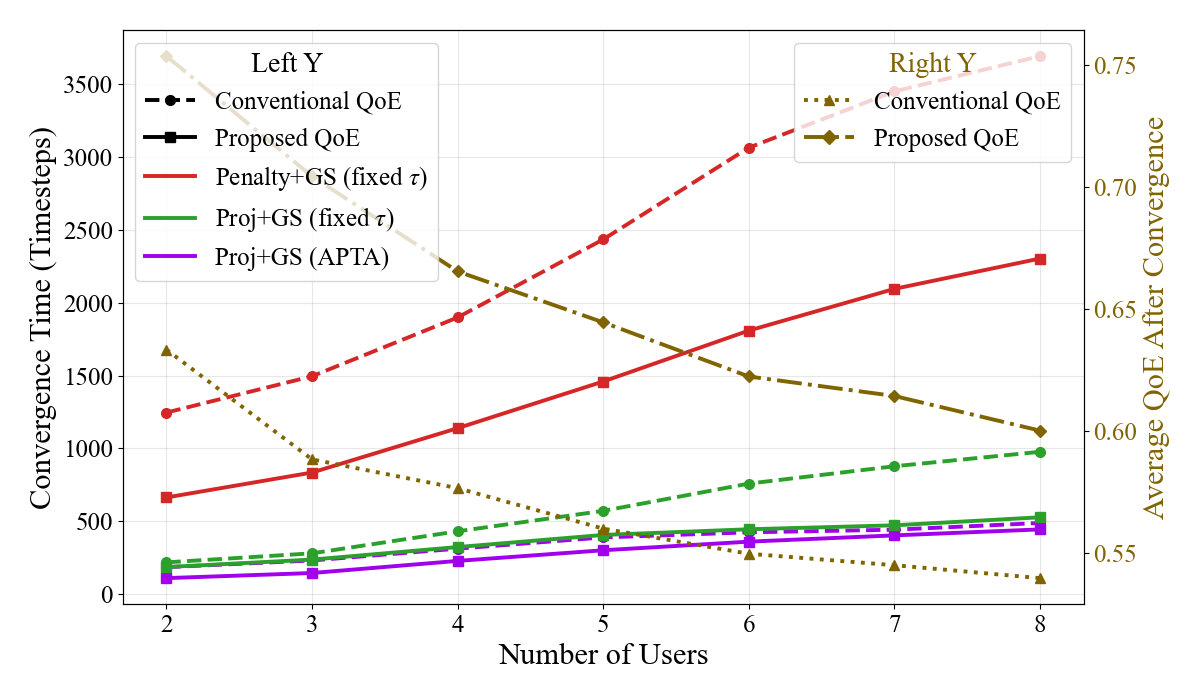}\vspace{-0.5em}
\caption{Curves for preference weight $w_n=0.2$.}
\label{fig:qoe_eval_curves}
\end{subfigure}\vspace{-0.5em}
\caption{Impact of the number of users on convergence time and average QoE after convergence for different action-handling strategies.}\vspace{-1em}
\label{fig:users_convergence_qoe}
\end{figure}

Fig. \ref{fig:users_convergence_qoe} plots the convergence time and achieved average QoE versus the number of users, $N$, for different action-handling strategies. First, when $N$ increases, all strategies exhibit monotonically increasing convergence time and decreasing average QoE, driven by the intensified competition for power and subcarrier resources. Moreover, the proposed QoE achieves improved convergence behavior and more effective learning outcomes compared with the conventional QoE, as it more effectively promotes the joint optimization of semantic rate and similarity. Furthermore, projection-based strategies exhibit significantly better scalability than penalty-based strategies, especially for large $N$, and Proj+GS (APTA) consistently achieves the fastest convergence due to its adaptive temperature control. As $N$ increases, projection-based strategies maintain stable convergence with a relatively slow growth in convergence time, while Penalty+GS (fixed $\tau$) exhibits rapidly increasing convergence time and unstable behavior. This further highlights the advantage of projection-based strategies in providing more stable and less noisy learning. 
In addition, we note that the complexities of Penalty+GS (fixed $\tau$) and Proj+GS (fixed $\tau$) are $\mathcal{O}(\mathcal{C}_{\mathrm{RL}}+NM)$ and $\mathcal{O}(\mathcal{C}_{\mathrm{RL}}+NM+N^2+N\log N)$, respectively, where $\mathcal{C}_{\mathrm{RL}}$ is the computational cost of one MADDPG training update. It indicates that Proj+GS (APTA) has the same asymptotic complexity as Proj+GS (fixed $\tau$), demonstrating that APTA improves convergence efficiency without increasing the asymptotic computational complexity.

% \textcolor{red}{Fig.~\ref{fig:users_convergence_qoe} also evaluates the computational efficiency in terms of the number of timesteps required for convergence. We further analyze the per-timestep computational complexity of the considered schemes. We denote $\mathcal{C}_{\mathrm{RL}}$ as the computational cost of one MADDPG training update. The complexities of Heuristic and Penalty+GS (fixed $\tau$) are $\mathcal{O}(N)$ and $\mathcal{O}(\mathcal{C}_{\mathrm{RL}}+NM)$, respectively, where $\mathcal{O}(\cdot)$ denotes the computational complexity. The power projection requires at most $N$ redistribution rounds, while the subcarrier repair requires sorting the fractional allocations. Therefore, the complexity of Proj+GS (fixed $\tau$) is $\mathcal{O}(\mathcal{C}_{\mathrm{RL}}+NM+N^2+N\log N)$. Since APTA introduces only constant-complexity scalar updates, Proj+GS (APTA) has the same asymptotic complexity as Proj+GS (fixed $\tau$).}

% \begin{figure}[!t]
% \centering
% \includegraphics[width=0.75\linewidth]{Pictures/users_convergence_qoe.png}
% \vspace{-0.5em}
% \caption{Impact of the number of users on convergence time and average QoE after convergence for different action-handling strategies.}\vspace{-1.8em}
% \label{fig:users_convergence_qoe}
% \end{figure}

\section{Conclusions}

This paper proposed a QoE-driven hybrid resource allocation framework for multi-user downlink OFDMA semantic communication systems. A non-compensatory multiplicative QoE was introduced to capture the joint performance of semantic rate and similarity, as reflected by the fact that the QoE vanishes if either the semantic similarity or rate reduces to zero, while remaining non-decreasing in both dimensions and allowing flexible preference control, leading to a utility more aligned with semantic service objectives. To solve the resulting mixed continuous-discrete problem under coupled resource constraints, we proposed a projection-based MADDPG framework with GS relaxation and APTA. Numerical results demonstrated that our proposed Proj+GS (APTA) scheme achieves faster convergence, improved stability, and higher QoE compared with conventional benchmarks. Specifically, projection-based constraint handling stabilizes learning by ensuring feasibility, while APTA improves the discrete decision learning efficiency by balancing exploration and discretization.

\section*{Acknowledgment}

This work was funded by the Australian Research Council Discovery Project DP260104319.

\end{document}